\documentclass[preprint2]{aastex}

\shorttitle{Radio Properties of the Supernova Remnant \objectname[]{N157B}}
\shortauthors{Lazendic et al.}

\begin{document}
 
\title{Radio Properties of the Supernova Remnant \objectname[]{N157B} }

\author{J S Lazendic\altaffilmark{1}}
\affil{Centre for Astronomy, University of Western Sydney, PO Box 10,
Kingswood NSW 2747, Australia \\
Australia Telescope National Facility, CSIRO, PO Box 76, Epping NSW 1710, 
Australia}
 
\author{J R  Dickel}
\affil{Astronomy Department, University of Illinois at Urbana-Champaign, 1002 West Green Street, Urbana IL 61801, USA}

\author{R F Haynes}
\affil{Australia Telescope National Facility, CSIRO, PO Box 76, Epping NSW 1710, Australia}

\and

\author{P A Jones and G L White}
\affil{Centre for Astronomy, University of Western Sydney, PO Box 10, Kingswood NSW 2747, Australia}

\altaffiltext{1}{Current address: Astrophysics Department, School of Physics A29, University of 
Sydney, NSW 2006, Australia}

\begin{abstract}
A new investigation of the supernova remnant (SNR) \objectname[]{N157B} was carried out with
the Australia Telescope Compact Array. Radio continuum images of the entire
\objectname[]{30 Doradus} region have been made at 3.5 and 6\,cm wavelength with a resolution of 
$\sim$ 2\arcsec. These data allow a high resolution study of the
spectral index distribution and polarization properties of both N157B and the nearby \objectname[]{30 Doradus} nebula (the latter will be reported in a subsequent paper). N157B is an extended Crab--type SNR which may be beginning
the transition to a composite remnant.  There is
little apparent fine structure and the brightest radio region is
several parsecs from the probable position of the X--ray pulsar.  The
SNR has a radio spectral index of $-$0.19 and is significantly polarized 
at 3.5\,cm but not at longer wavelengths.
\end{abstract}


\section{Introduction}

\objectname[]{N157B} \citep{hen56} (also 30 Dor B or SNR 0538--691)
is a supernova remnant located near the edge of the 
\objectname[]{30 Doradus} nebula \citep{bod01} in the \objectname[]{Large Magellanic Cloud} (LMC). The
\objectname[]{30 Doradus} complex is the nearest extragalactic giant \ion{H}{2} region and an
active star--forming region \citep{chu92}. The
nebula and the SNR are in the same interstellar region of the LMC as
\objectname[]{SN1987A} and other SNRs, among which is \objectname[]{SNR 0540--693}, a
composite remnant with a pulsar and a plerion component \citep{man93}. The SNR nature of
\objectname[]{N157B} was first determined by its excess of radio over optical
emission \citep{lem68}, and confirmed by the
detection of [\ion{S}{2}] lines in the optical spectrum \citep{dan81}.
   
The Crab nebula represents the class of pleronic supernova 
remnants; it contains a pulsar as a central energy source and a 
well--organised magnetic field indicated by strong linear polarization
at radio wavelengths and relatively flat radio spectrum \citep{wei83}. A number of detected SNRs have radio 
characteristics similar to the \objectname[]{Crab nebula}, yet the pulsar stimulation 
of the synchrotron nebula is confirmed for just a few of them 
\citep{sew83,sew89,wei88}.  \objectname[]{N157B} is classified as a Crab--type remnant because of  
its centre--filled morphology, flat radio spectrum
  \citep{mil78,mil80}, and 
strong soft X--ray emission \citep{lon79}.

Recently a 16\,ms pulsar detected only in X--rays 
 \citep{mar98,got96,wan98}  has been associated with
\objectname[]{N157B}. Previous observations of \objectname[]{N157B} at radio wavelength have not
detected any polarized emission  \citep{dic94}. However, our current investigation at a shorter wavelength
shows a significant level of polarization. All of the data together
confirm the classification of \objectname[]{N157B} as a Crab--type SNR.

The presence of \objectname[]{N157B} in a relatively dense and complex interstellar 
environment together with the younger SN1987A and the presumably older 
SNR 0540--693, offers a unique opportunity to study the evolution of 
Crab--type SNRs in such areas.  The best Galactic examples, the Crab 
and 3C58, both appear to be in exceptionally low--density environments 
 \citep{fes97}.  

\section{Observations and data reduction}

\subsection{General procedures}

The first obervations at 6\,cm were carried out with the 
Australia Telescope Compact Array\footnote{The Australia Telescope is funded
by the Commonwealth of Australia for operation as a National Facility managed
by CSIRO} (ATCA) in the period January -- July 1993. Additional observations at
3.5 and 6\,cm were obtained during May 1998 -- June 1999. The six antennas were 
used in several different array configurations giving baselines with the range 
from 31\,m to 6\,km to provide good sampling of the observed region. Details 
of the ATCA can be found in \citet{fra92}. The relevant observing parameters 
are given in Table \ref{tbl-obs}.

\begin{deluxetable}{llll}
\scriptsize
\tablecaption{Observational parameters. \label{tbl-obs}}
\tablewidth{0pt}
\tablehead{
\colhead{Parameter} & \colhead{4.4\,GHz} & \colhead{4.8\,GHz} & \colhead{8.6\,GHz}}

\startdata

Date .................................. & 1998 Aug 27 & 1993 Jan 16, 19, 29 &  1998 May 1, 31, Aug 27 \\                                               
                                           & 1998 Sep 26, Oct 2& 1993 Jul 23 &  1998 Sep 26, Oct 222\\
                                            &  1999 Jun 24 &    & 1999 Jun 24\\
Configuations .....................& 6C, A, 1.5D, 375 & 750B, C, A, D &
                                     750A, E, 6C, A, 1.5D, 375\\
                                              
Field Center (J2000) ..........& $05^h37^m50^s$\tablenotemark{a}& $05^{h}38^{m}45^{s}$  &
$05^{h}37^{m}50^{s}$\tablenotemark{a}      \\
                                & $-69\arcdeg10\arcmin00\arcsec$\tablenotemark{a} & $-69\arcdeg06\arcmin00\arcsec$ & 
                                 $-69\arcdeg10\arcmin00\arcsec$\tablenotemark{a}   \\
Total bandwith (MHz) ......& 128    & 128    & 128  \\
No. of frequency channels ..& 32     & 32    &  32   \\
FWHM of primary beam ...& 10\arcmin & 10\arcmin & 5\arcmin\\

\enddata
\tablenotetext{a}{field no.\,15 of a larger mosaic}
\end{deluxetable}

The primary calibrator used in all observations was PKS B1934--638 with assumed 
flux densities of $\sim$6 and 2.78\,Jy at 6 and 3.5\,cm respectively. PKS 
B0454--810 was used as secondary calibration source.

The observations in 1998 -- 1999 were done in mosaic mode with 20 pointing centres to
cover the whole \objectname[]{30 Doradus} region. The observing pattern was such that
the pointing nearest to the SNR (field no.\,15 -- see Table\,\ref{tbl-obs}) 
was observed after every
five fields to improve signal--to--noise on this object. The data were
taken simultaneously in 32\,$\times$\,4--MHz frequency channels of which the 26 central
channels are used in the imaging procedure. They have been averaged
down to 13\,$\times$\,8--MHz channels by Hanning smoothing and then to a single
channel during imaging. The data were gridded and deconvolved according
to the procedure for mosaic data sets \citep{sau97}.
 Separate images of field no.\,15 alone were also produced.

The 3.5--cm images of the SNR were made by combining 7 fields 
(including no.\,15)
which had the pointing centre closest to the SNR. The deconvolution of
all 7 fields was done together, which improves the deconvolution
process but introduces some errors in the model of the primary
beam. The edges of the mosaiced region have low sensitivity so the
noise level across the mosaic image will vary. In the imaging process
the primary beam attenuation will not be completely corrected, because
that correction is constrained by the noise in the individual
pointings, which can differ \citep{sau97}. In this case, the procedure limits 
the dynamic range of the
final image to 100:1. We also made an image of field
no.\,15 alone. This  was cleaned and corrected for the pattern of 
the primary beam. The peak brightness in both images is very 
close ($\sim$7\,mJy/beam), but the single field image has lower 
signal--to--noise ratio over the entire field.

The image at 6\,cm was constructed by combining four data sets at 
4.8\,GHz and pointing no.\,15 from 4.4\,GHz data 
(see Table~\ref{tbl-obs}).
Thus, in addition to the mosaicing procedure explained above, the 
multi--frequency synthesis technique \citep{sau97} was also used 
to produce the final image of the region. As in the previous case, an image
of field no.\,15 alone was made as well. 

\subsection{Polarization}

The interferometer produces an instrumental polarization which
increases outward from the center of the primary beam pattern of the
individual telescopes. In fields with the pointing centre away from the
SNR for which an unpolarized source (30 Doradus) is at the half--power
point, we found fractional polarizations up to $\sim$10\% with electric field  vectors radially oriented with respect to the direction of the
beam centre. This value of fractional polarization in the surrounding
fields confirms that off--centre instrumental polarization is indeed
strong.

Therefore, for the polarimetry we used only field no.\,15 for which
\objectname[]{N157B} is in the center of the beam and the instrumental polarization is
low (0.3\%). The images of the $Q$ and $U$ Stokes parameters were made
in a similar way as the $I$ image and combined to produce the image of
polarized intensity. To create such an image we used the MIRIAD task IMPOL 
which statistically corrects the polarized intensity for Ricean bias 
 \citep{sau97}. Noise in both the $Q$ and $U$ 
polarization images was below the 5--$\sigma$ noise level of the total intensity image and CLEANing was not necessary.  

The polarized intensity images of the SNR at 6\,cm were made using only field
 no.\,15
with the same procedure as for the single--field image explained
above. The 1993 positions were not directly toward
N157B and so were not included in the polarization analysis because
of the off--axis effects disscussed above. The fractional polarization in the 
direction of \objectname[]{N157B} was found
to be $\sim$3\%, near the instrumental value, thus confirming
only that \objectname[]{N157B} is not highly polarized at this wavelength. 

\section{Results}

\subsection{3.5\,cm image}

\begin{figure}
\plotone{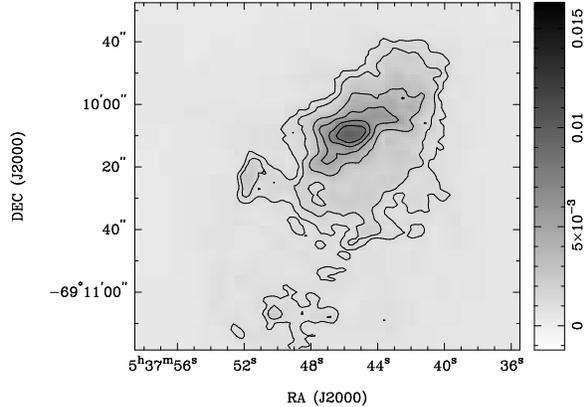}
\caption{Radio continuum grayscale and contour image of N157B at 3.5\,cm 
with HPBW of 2\farcs67 $\times$ 1\farcs80. The contour levels are: 1.1, 1.6, 3.3, 
4.9, 6.5, 8.2 and 9.8\,mJy\,beam$^{-1}$.} 
\label{fig-snr3cm}
\end{figure}

The high resolution total intensity image at 3.5\,cm of the region
around \objectname[]{N157B} was made using all six available configurations; it
has a Gaussian beam of HPBW $\sim$2\arcsec\ (see Fig.\ref{fig-snr3cm}). It shows 
reasonably uniform brightness
distribution and the center--filled morphology characteristic of
Crab--type SNRs. We note that the bright peak is resolved.  It lies
about $12\arcsec$ (4\,pc at the 50\,kpc distance to the LMC) to the northwest of 
the X--ray peak, which is presumably
the location of the pulsar \citep{wan98}. Also, the radio peak is not centered on the full SNR which extends
well to the south with dimensions of about $120\arcsec\times70\arcsec$ or 
30\,pc$\times$18\,pc. Slices along the major axis of the remnant at a
position angle of about 150$\arcdeg$  
show continuous emission throughout the total extent with a slight
decrease near a declination of $-69\arcdeg10\arcmin50\arcsec$ and then a rise again to the
south. Perhaps part of an outer shell 
may be forming around the plerion. The size given includes this outer 
emission. 

This is the most sensitive high--resolution radio image of this object
now available, yet no direct evidence of the pulsar can be seen. This 
is not surprising, however, as 3.5\,cm is a very short wavelength for 
pulsar emission which usually peaks at meter wavelengths. The pulsar has 
not been found at any radio wavelength.

\begin{figure}
\plotone{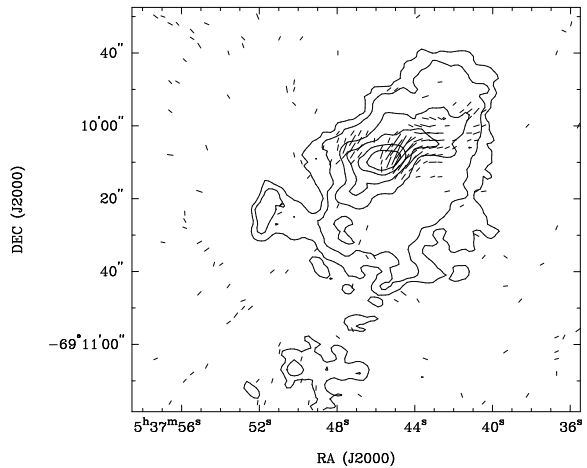}
\caption{Image of the total intensity of N157B at 3.5\,cm with 
{\bf E}--vectors superimposed.  A vector length of $1\arcsec$ 
  represents a linearly polarized intensity of 1.7\,mJy\,beam$^{-1}$. The contour 
  levels are: 1.1, 1.6, 3.3, 4.9, 6.5, 8.2 and 9.8\,mJy\,beam$^{-1}$.}
\label{fig-pol3cm}
\end{figure}

We found strong linear polarization in the direction towards the SNR. The
electric field vectors are superimposed on continuum contours in 
Fig.\ref{fig-pol3cm}. The mean fractional polarization is $\sim$10\%, 
calculated by dividing the polarized intensity image of the source by
the total intensity image (both made with field no.\,15 only). The total 
intensity values were truncated at the 3--$\sigma$ level which is 
$\sim$5\,mJy\,beam$^{-1}$ before division, but all (bias corrected) polarized 
intensities within the source boundaries given by the total intensity
cutoff were used.  In the vector display, the vectors were truncated at
the 3--$\sigma$ level. We note that a polarization map using seven
mosaic positions looked similar but was noisier because of the 
few percent radial instrumental polarization toward \objectname[]{N157B} from the 
surrounding fields. 

The electric vector alignment is fairly uniform but the polarized 
intensity is surprisingly patchy.  The highest fractional polarization
is 15.4\% towards $05^h37^m44\fs5$ and $-69\arcdeg10\arcmin07\farcs5$\footnote{all position are in J2000}, whereas 
at the  position of the brightest total intensity $05^h37^m41\fs5$ and 
$-69\arcdeg09\arcmin31\farcs0$ it is 5\%, which is the lowest fractional
polarization (excluding blanked regions). At the position of the X--ray
peak at $05^h37^m47\fs5$ and $-69\arcdeg10\arcmin17\farcs5$ the polarized
intensity is 7\%.

\subsection{6\,cm image}

\begin{figure}
\plotone{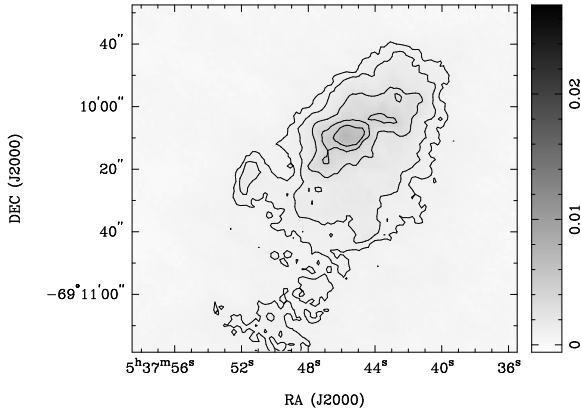}
\caption{Radio continuum grayscale and contour image of N157B at 6\,cm 
with HPBW of 1\farcs79 $\times$ 1\farcs74. The contour levels are: 0.9, 1.3, 2.4, 4.0, 
5.4, 8.1, 10.8 and 13.5\,mJy\,beam$^{-1}$.}  
\label{fig-snr6cm}
\end{figure}

The total intensity image of \objectname[]{N157B} at 6\,cm (shown in Fig.\ref{fig-snr6cm}) was
 made using eight available configurations of the Compact Array 
 (see Table\,\ref{tbl-obs}) and 
 also has the synthesized beam of HPBW $\sim$2\arcsec. 
The fractional polarization towards \objectname[]{N157B} (calculated
using only field no.\,15) is only $\sim$3\%, which is close to the
noise level.  We therefore detect no reliable polarization at 6\,cm and
conclude that there must be strong Faraday depolarization between 3.5
and 6\,cm.  We note that at 6\,cm the distribution of the closest antenna
spacings represents lower spatial frequencies than those seen at 3.5\,cm. If 
the polarization is all on a fine spatial 
scale we may find a lower value of the polarization at 6\,cm.
Inspection of Fig.\ref{fig-pol3cm}, however,
shows that while there is fine structure a significant part of the
3.5--cm polarization is uniform on scales of over $10\arcsec$ and so should
appear at 6\,cm. Depolarization effects are obviously present.

\section{Discussion}

\objectname[]{N157B} is very close to \objectname[]{30 Doradus} and it is also embedded in an \ion{H}{2} 
region around the OB association \objectname[]{LH99} \citep{luc70}.
\citet{chu92} concluded that the southern part of the SNR
is obscured and possibly interacts with the dark cloud within the \ion{H}{2} region.
With our high radio sensitivity we can see that the SNR does extend well to 
the south of the bright northern component. With a size of 30\,pc$\times$18\,pc, 
it is the largest Crab--type SNR known.  The feature at $05^h37^m52^s$ and 
$-69\arcdeg10\arcmin22\arcsec$ and the
elongated structure centered at about $05^h37^m50^s$ and $-69\arcdeg11\arcmin10\arcsec
$ might indicate the beginning of a transition into a composite remnant 
like SNR 0540--693.  The dense interstellar environment of the 
\objectname[]{30 Doradus} region may be conducive to the formation of a shell and the irregular 
structure of \objectname[]{N157B} may be accounted for by a complex density distribution.

\citet{chu92} found that \objectname[]{N157B} contributes up
to 80\% of the flux density in this region at 0.843 GHz , the rest
coming from the surrounding \ion{H}{2} region. They suggested that the flat
radio spectral index ($\alpha$) of $-$0.1 ($S_{\nu} \approx \nu^{\alpha}$) indicated from
earlier measurements \citep{mil80} should be
characteristic of the SNR itself. However, they also found that the
optical emission lines from \objectname[]{N157B} originate in conditions associated
with  blast waves and they therefore concluded that the corresponding
non--thermal radio emission should not have the flat spectrum typically
associated with pulsar stimulation. Our results below differ somewhat
from the previously published values derived from low resolution data but still
show that N157B has a flat radio spectrum.
 
Data used for determining the radio spectrum of \objectname[]{N157B} are listed in Table\,
\ref{tbl-spind}. To obtain a
consistent background and source extent, the images were all convolved
to the $43\arcsec\times45\arcsec$ resolution of the MOST image and
the integrated flux density calculated over the same area in each
convolved map. A background intensity around the source was determined
from the average of several areas around the SNR but far enough from 30
Doradus to avoid contamination.  

\begin{deluxetable}{lll}
\footnotesize
\tablecaption{Flux Densities of N157B. \label{tbl-spind}}
\tablewidth{0pt}
\tablehead{
\colhead{Frequency (GHz)} & \colhead{Flux Density (Jy)} & \colhead{References}}

\startdata
0.843  & $2.86 \pm 0.24$ & Anne Green (1997, private comunication) \\
1.400  & $2.64 \pm 0.16$ & Data of \citet{meb97} \\ 
2.378  & $2.36 \pm 0.14$ & Data of \citet{dic94} \\
4.740  & $2.20 \pm 0.10$ & Data of this paper \\
8.600  & $1.82 \pm 0.12$ & Data of this paper \\

\enddata
\end{deluxetable}

\begin{figure}
\plotone{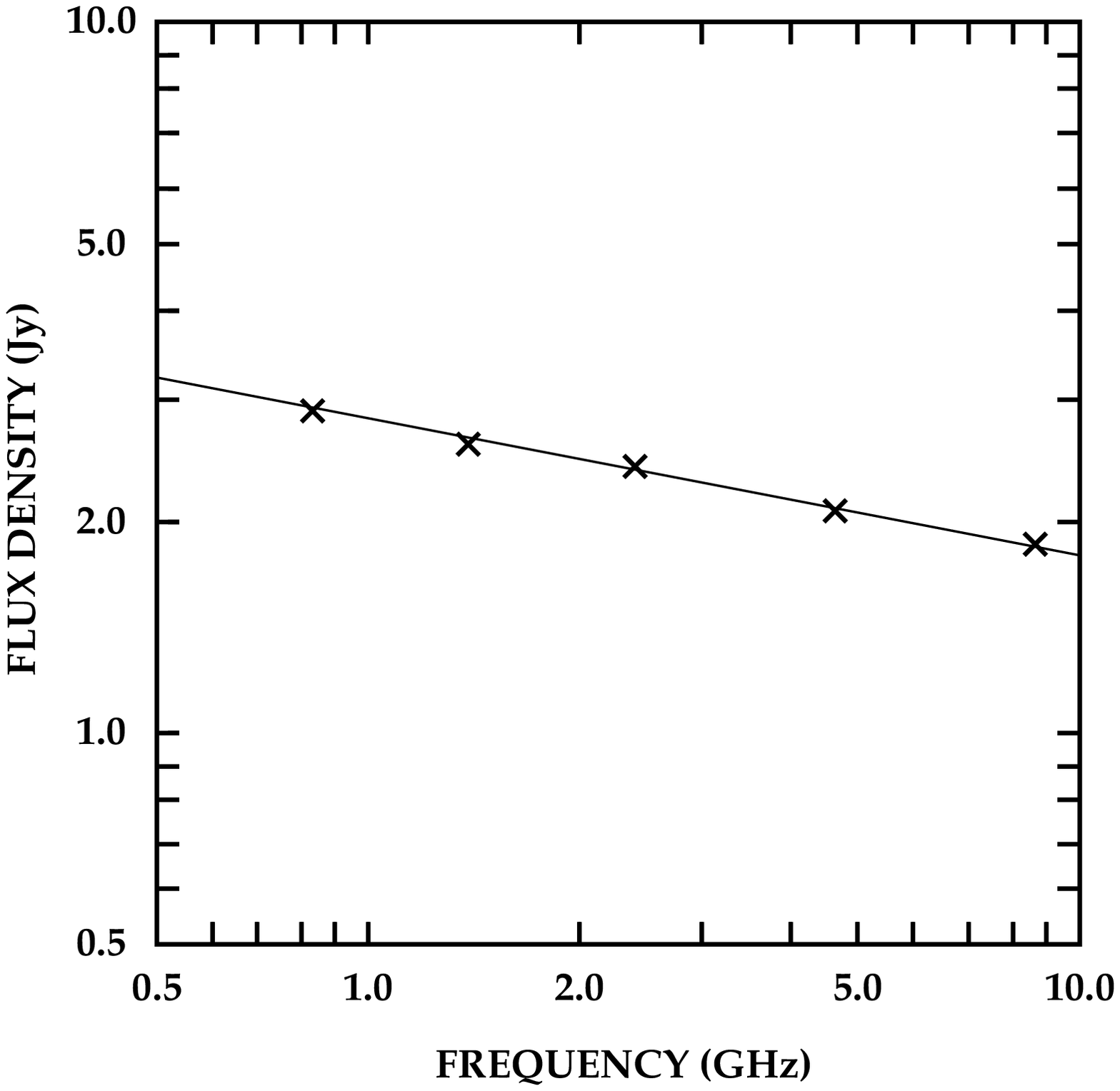}
\caption{Radio spectrum of N157B. The slope of the fitted line is $-$0.19.}
\label{fig-spind}
\end{figure}

The resultant integrated flux densities listed in Table \ref{tbl-spind}
give a mean spectral index for \objectname[]{N157B}
of $-$0.19 (see Fig.\ref{fig-spind}). Brightness--brightness
plots also show that the spectrum does not vary significantly across
the whole extent of the SNR. This argues that the entire SNR is pulsar--powered.

In the X--ray band, however, the spectrum of \objectname[]{N157B} differs slightly
relative to other such remnants. It has a power--law energy slope
of $-$1.5 which is steeper than those of other Crab-like remnants with
values of about $-$1.0 \citep{wan98}. The
X--ray and entire spectrum of \objectname[]{N157B} also strikingly resemble that of SNR
0540--693, leaving no doubt of the pleronic origin of \objectname[]{N157B}. The SNR
0540--693 contains a pulsar and is a composite remnant, sharing the
characteristics of pure shell remnants and pure plerions
\citep{sew89,man93}. This latter
composite SNR is often called a Crab--type remnant, because the core is
Crab--like and the faint shell component was only investigated
later. This further similarity between \objectname[]{N157B} and 0540--693 is 
more evidence that \objectname[]{N157B} is beginning the transition to a composite remnant.    

The polarization properties of the \objectname[]{30 Doradus} region were previously
investigated by \citet{dic94} at longer
wavelengths. They used ATCA observations at 13\,cm but concluded that
no polarized emission was detected in any region of \objectname[]{N157B} above the
level of instrumental polarization (less than 2\%). We report a
significant level of polarized emission at 3.5\,cm from \objectname[]{N157B} for the
first time. This relatively high level of 10\% is typical for
a Crab--type remnant. The data at 6\,cm give a marginal detection which we
consider an upper limit of 3\%
fractional polarization at the peak of \objectname[]{N157B}. This low level of
polarized emission at this wavelength implies very strong internal 
depolarization. The observed polarization could also be reduced by 
the thermal emission from the associated \ion{H}{2} region around \objectname[]{LH99} which
may be intertwined with the SNR so that its thermal electrons may cause
some of the Faraday depolarization as well. 

To estimate the internal rotation  necessary to cause the observed 
depolarization, we use a slightly modified version of the relation 
given by \citet{bur66}:

\begin{equation}
P(\lambda_2)/P(\lambda_1) = sin(\delta)/\delta
\end{equation}

\noindent where $P$ is the fractional polarization at $\lambda_{2}$, the longer wavelength, 
and $\lambda_{1}$, the shorter wavelength; $\delta$ is the rotation in
position angle of the measured electric vector between the two
wavelengths.  We will adopt the possible detection of 3\% polarization
at 6\,cm and so use a rough mean value for $P(\lambda_{2})/P(\lambda_{1})$ of
0.3 which requires an average rotation of 135$\arcdeg$.  Then using the
standard formula for Faraday rotation we can estimate the line--of--sight
magnetic field strength:

\begin{equation}
\delta = 7.9 \times 10^5 \lambda^2 \int N_e \vec{B} \dot{\times} \vec{dl}
\end{equation}

\noindent where $\delta$ is in radians, $\lambda$ is in meters, $N_{e}$ is the electron
density in cm$^{-3}$, $B$ is the magnetic field strength in Gauss, and $l$ is
the path length through the SNR in pc.  Because the Faraday rotation
increases progressively along the path, we must actually find the adopted
rotation along half the
depth along the line of sight.  We do not, of course, know this depth
directly but choose its value to be approximately equal to the short 
observed axis of the SNR or a half depth of 10\,pc.  From an analysis
of ROSAT and ASCA X--ray spectra, \citet{wan98} derived an
electron density of $0.6f^{-0.5}$\,cm$^{-3}$, where $f$ is the filling factor of the
object.  Because Crab--type SNRs should be reasonably uniform inside, we
adopt a value for $f$ of about 0.5 to get a mean density of about 2\,cm$^{-3}$. 
 These numbers give a line--of--sight magnetic field strength of
30\,$\mu$Gauss.  Most Crab--type SNRs have rather uniform magnetic fields;
over the limited region where it can be observed, \objectname[]{N157B} shows more
variation than most.  It may be that the field is directed mainly
toward us, which would reduce the observed polarization at all
wavelengths and also give a large Faraday rotation.  No correction has been
made for an unknown field orientation which would tend to
increase the field strength in the remnant.  We should emphasize that 
this result is only an order
of magnitude estimate.  Neglected factors include the uncertainties and
probably variations in the depolarization, magnetic field reversals,
density uncertainties and variations.  For example, significant
clumping can change the apparent integrated electron density because
the X--ray emission depends upon the density squared but the Faraday
rotation depends linearly on the density and the sum of squares does
not equal the square of sums. 

If we know the magnetic field strength, we can then determine the energy in 
relativistic electrons from the intensity of the synchrotron
radiation.  \citet{gin65} give the
approximate formula: 

\begin{equation}
E_e = \Sigma_\nu \times R^2 \times A \times area \times pathlength \times B^{-3/2}
\end{equation}

\noindent where $E_{e}$ is the total energy in relativistic electrons, $\Sigma_{\nu}$ is
the surface brightness of the remnant at a given observing frequncy $\nu$, 
$R$ is its distance, and $A$ is a
parameter that depends on the frequency, the spectral
index and the adopted cutoff frequencies for the synchrotron
radiation.  There appears to be a break in the spectrum of the
non--thermal emission between the radio and X--ray \citep{wan98}  but we do not know specifically where it occurs.  We
shall choose an upper frequency of $10^{12}$\,Hz and a lower one of $10^{7}$\,Hz, with a spectral index of $-$0.19, $A = 3.7 \times 10^{19}$ at 8.6\,GHz where the
flux density is 1.82\,Jy.  With the sizes and magnetic field strength
given above, we find that the mean relativistic electron energy density
is $4\times10^{-10}$\,ergs\,cm$^{-3}$.  This is about 10 times the magnetic energy
density.  While within the errors, we cannot definitely state that the
values are different. It is interesting to note that in the three
objects for which we have now measured the relative energies by this
method, each has between 3 and 10 times greater relativistic electron 
energy than magnetic energy (Kepler's SNR \citet{mat84} and N23 SNR \citet{dic98}). 

\citet{wan98} give an approximate temperature for the
thermal gas in \objectname[]{N157B} of 0.4 -- 0.7\,keV.  Choosing 0.6\,keV and the
approximate density of 2\,cm$^{-3}$ used above, the thermal energy density is
about $3\times10^{-9}$\,erg\,cm$^{-3}$.  Thus the thermal energy is about 10 times
that in relativistic electrons and 100 times that in the magnetic field.
Even in this Crab--type SNR the thermal energy dominates.

In summary, the filled structure and uniform spectral index of 
\objectname[]{N157B} are Crab--like but it also has some characteristics similar to
those of the composite SNR 0540--693.  Although the dense, irregular
surroundings may be partly responsible for the non--uniform brightness
and outline of this SNR particularly to the south, the overall
morphology and spectral index suggest that this SNR is just beginning
the transition from a pure Crab--like remnant to a composite remnant.
It will be interesting to see how it changes at radio, optical and
X--ray wavelengths in the future.

We have detected for the first time a significant level of linearly 
polarized emission in this object at 3.5\,cm wavelength. The 
encompassing H II region and clumpy internal structure within the SNR 
may cause sufficient Faraday rotation to reduce the polarization to 
below detectable levels at 6\,cm and make the large variations in 
the polarization observed at 3.5\,cm at different positions around the 
remnant.  Observations at higher frequencies, e.g. 23\,GHz, are needed  
to enable more accurate determination of the Faraday rotation.

\acknowledgments

We thank M. Costa, D. Milne and H. Dickel for participating in the observing. 
We thank L. Staveley-Smith for the ATCA image at 21\,cm and helpful comments.
 We thank B. Gaensler and A. Green for providing us with the MOST image. JRD 
 acknowledges support from the Campus Honours Program of the University of 
 Illinois at Urbana-Champaign. JSL acknowledges support from the Astronomy DRG and the 
 UWS Nepean Research Office. Work at UWS Nepean was also supported by the ARC.
  JSL thanks the ATNF for the use of their facilities, J. Whiteoak for useful 
  comments on polarization, N. Killeen and R. Sault for valuable discussions 
  on data reduction, and M. Filipovic.




\clearpage

\begin{thebibliography}{}

\bibitem[Bode(1801)]{bod01} Bode, J., 1801, `Algemeine Beschriebung und Nachweisung der Gestirne', Berlin: Beym Verfasser
\bibitem[Burn(1966)]{bur66} Burn, B. J., 1966, \mnras, 133, 67
\bibitem[Chu et al.(1992)]{chu92} Chu, Y.-H., Kennicutt Jr., R. C., Schommer, R. A. \& Laff, J., 1992, \aj, 103, 1545
\bibitem[Clarke et al.(1976)]{cla76} Clarke, J. N., Little, A. G. \& Mills, B. Y., 1976, Aust. J. Phys. Astrophys. Suppl., 40, 1
\bibitem[Danzinger et al.(1981)]{dan81} Danziger, I. J., Goss, W. M., Murdin, P., Clark, D. H. \& Boksenberg, A., 1981, \mnras, 195, 33
\bibitem[Dickel et al.(1994)]{dic94} Dickel, J. R., Milne, D. K., Kennicutt, R. C., Chu, Y.-H. \& Schommer, R. A., 1994, \aj, 107, 1067
\bibitem[Dickel \& Milne(1998)]{dic98} Dickel, J. R., Milne, D. K., 1998, \aj, 115, 1057
\bibitem[Fesen(1997)]{fes97} Fesen, R. A. 1997, presented in `10$^{51}$ Ergs,' SNR Workshop, Minneapolis MN, 1997
\bibitem[Frater et al.(1992)]{fra92} Frater, R. H., Brooks, J. W. \& Whiteoak, J. B., 1992, J. of Electrical \& Electronic Eng. Austr., 12, 103
\bibitem[Ginzburg \& Syrovatskii(1965)]{gin65} Ginzburg, V. L. \& Syrovatskii, S. I., 1965, \araa, 3, 297
\bibitem[Gotthelf \& Wang(1996)] {got96} Gotthelf, E. V. \& Wang, Q. D., 1996, `Roentgen strahlung from the Universe,' MPE Report 263, 255
\bibitem[Green(1997)]{gre97} Green, A., 1997, private communication
\bibitem[Green(1996)]{gre96} Green, D. A., 1996, `A catalogue of galactic supernova remnants', Cambridge: Mullard Radio Astronomy Observatory, 1996 August version
\bibitem[Haynes et al.(1991)]{hay91} Haynes, R. F., Klein, U., Wayte, S. R., Wielebinski, R., Murray, J. D., Bajaja, E., Meinert, D., Buczillowski, U. R., Harnett, J. I., Hunt, A. J., Wark, R. \& Sciacca, L., 1991, \aap, 252, 475
\bibitem[Henize(1956)]{hen56} Henize, K. G., 1956, \apjs, 2, 315
\bibitem[Le Marne(1968)]{lem68} Le Marne, A. E., 1960, \mnras, 139, 461
\bibitem[Long \& Helfand(1979)]{lon79} Long, K. S. \& Helfand, D. J., 1979, \apj, 234, L77
\bibitem[Lucke \& Hodge(1970)]{luc70} Lucke, P. B. \& Hodge, P. W., 1970, \aj, 75, 171
\bibitem[Manchester et al.(1993)]{man93} Manchester, R. N., Staveley-Smith, L. \& Kesteven, M. J., 1993, \apj, 411, 756
\bibitem[Marshall et al.(1998)]{mar98} Marshall, F. E., Gotthelf, E. V., Zhang, W., Middleditch, J. \& Wang, Q.D., 1998,  \apjl, 499, L179
\bibitem[Matsui et al.(1984)]{mat84} Matsui, Y., Long, K. S., Dickel, J. R. \& Greisen, E. W., 1984, \apj, 287, 295
\bibitem[Mebold et al.(1997)]{meb97} Mebold, U., Deusterberg, C., Dickey, J. M., Staveley-Smith, L. \& Kalberla, P., 1997, \apj, 490, 65
\bibitem[Milne et al.(1980)]{mil80} Milne, D. K., Caswell, J. L. \& Haynes, R. F., 1980, \mnras, 191, 496
\bibitem[Mills et al.(1978)]{mil78} Mills, B. Y., Turtle, A. J. \& Watkinson, A., 1978 \mnras, 185, 263 
\bibitem[Sault \& Killeen(1997)]{sau97} Sault, R. J. \& Killeen, N., 1997, `Miriad Users Guide,' ATNF Publication
\bibitem[Seward(1983)]{sew83} Seward, F. D., 1983, in `Supernova Remnants and their X-Ray Emission', IAU Symp. 101, (eds.) J. Danziger \& P. Gorenstain, (Dordrecht, Kluwer), 405
\bibitem[Seward(1989)]{sew89} Seward, F. D., 1989, Space Science Reviews, 49, 385
\bibitem[Wang \& Gotthelf(1998)]{wan98} Wang, Q. D. \& Gotthelf, E. V., 1998, \apj, 494, 623
\bibitem[Weiler(1983)]{wei83} Weiler, K. W., 1983, in `Supernova Remnants and their X-Ray Emission', IAU Symp. 101, (eds.) J. Danziger \& P. Gorenstain, (Dordrecht, Kluwer), p. 299
\bibitem[Weiler(1988)]{wei88} Weiler, K. W. \& Sramek, R. A., 1988, \araa, 26, 295 
 
\end{thebibliography}
\end{document}